\documentstyle[12pt]{article}

\begin{document}

\thispagestyle{empty} 
\renewcommand{\thefootnote}{\fnsymbol{footnote}}

\begin{center}
\vspace{1truecm} {\large \textbf{On Different Formulations of Chiral
Bosons}}

\vspace{1cm} \textbf{R. Manvelyan$^{1}$ \footnote{E-mail:
manvel@physik.uni-kl.de} \footnote{Alexander von Humboldt Fellow,%
\newline
\,\,\,\,\,\,\,\,\,\, On leave from Yerevan Physics Istitute},}

\textbf{R. Mkrtchyan$^{2}$ \footnote{E-mail: mrl@amsun.yerphi.am},
H.J.W. M\"{u}ller--Kirsten$^{1}$ \footnote{E-mail:
mueller1@physik.uni-kl.de}}\vspace{1cm}

$^{1}$\textit{Department of Physics, University of Kaiserslautern,}

\textit{P.\ O.\ Box 3049, D 67653 Kaiserslautern, Germany} \vspace{1cm}

$^{2}$\textit{Theoretical Physics Department,} \textit{Yerevan Physics
Institute,}

\textit{Alikhanian Br. St.2, Yerevan, 375036 Armenia }
\end{center}

\vspace{1cm}

\begin{abstract}
It is shown, that recently constructed PST Lagrangians for chiral
supergravities follow directly from earlier Kavalov--Mkrtchyan 
Lagrangians by an Ansatz for the $\theta $ tensor by expressing this 
in terms of the PST scalar. The susy algebra which
included earlier $\alpha$--symmetry in the commutator of supersymmetry
transformations, is now shown to include both PST symmetries, which
arise from the single $\alpha$--symmetry term. The Lagrangian for the $5$-brane
is not described by this correspondence, and probably can be obtained from more
general Lagrangians, posessing $\alpha$--symmetry. 
\end{abstract}

\renewcommand{\thefootnote}{\arabic{footnote}} \setcounter{footnote}0

{\smallskip \pagebreak }

\section{Introduction}

Among fields appearing in  modern higher-dimensional theories are
the (anti)--self--dual tensor fields, describing the representations of
corresponding little groups with the constraint of (anti)--self--duality. Such
tensors have a rank, corresponding to dimensionality and signature of
space--time, and satisfy first--order field equations -- the condition of
self--duality of their field strength. In space--time with one time
dimension the duality condition is possible in dimensions $2,6,10$, etc. 
In $2$ dimensions the fields  are the chiral scalars, used 
e.g. in heterotic string theories \cite{HetStr}. 
In dimensions $6$ and $10$ the first theories considered, including as
a necessary part a self--dual (or anti--self--dual) tensor, were supergravities 
\cite{Sugra2B}, particularly $10$ dimensional  ${\cal N}=2b$ supergravity - one of two maximal
supergravities (another one is $11d$ ${\cal N}=1$ supergravity \cite{11dSugra}),
dimensional reduction of which to $4d$ gives maximally extended ${\cal N}=8$
supergravity \cite{ExtSugra}. Recently another theory with self--dual field
attracted much attention -- the six--dimensional theory, known as $5$-brane theory \cite
{5brane}\cite{Inter}, which is an important feature of M--theory. The unique
property of these fields is their contribution to the gravitational anomaly \cite
{AGW}, the only contribution coming from bosonic fields. This property is
intimately connected with the fact, that it is impossible to write down
the Lagrangian for such a field in the usual way -- and correspondingly the
regularization of quantum theory meets a difficulty, which eventually leads
to the appearance of the anomaly.

The problem of construction of Lagrangians for dual tensors, and eventually
Lagrangians for theories mentioned above was addressed in a number of
papers. Several methods were suggested for the construction of Lagrangians,
among which are that of Siegel \cite{Siegel}, the infinite
auxiliary field method \cite{Infinfield}, and the recently invented PST
formalism \cite{PST}. The generalization of the first one was used in the
papers of Kavalov and Mkrtchyan about ten years ago\cite{KM}, in which 
Lagrangians for all chiral supergravities were constructed for the first time.
Recently the equivalence of the PST and infinite auxiliary field methods was
claimed \cite{equiv}, and the Lagrangians for supergravities in the PST
formalism were constructed \cite{New2BSugra}. The aim of the present paper
is to show, that Lagrangians of ref. \cite{PST} follow directly from those of 
ref. \cite{KM} from an ansatz for the Lagrangian multiplier $\theta_{\mu \nu \lambda
,\sigma \rho \delta }^{++}$, and that the symmetry algebra, which includes the
so--called $\alpha$--symmetry, transforms into the algebra, containing specific
PST symmetries, the origin of these terms being exactly the $\alpha$--symmetry
terms. The correspondence with the method of ref. \cite{Infinfield} can also be
established. In the following the detailed proof is given of the equivalence of
the Lagrangians and the relations between the symmetry transformations. Section 
$2$ is devoted to definitions and the description of properties of self--dual
tensors. Sections $3$ and $4$ present the  Lagrangians of Kavalov and
Mkrtchyan, as well as $\alpha$--symmetry and the PST formulation. The ansatz for
the transformation of one theory into another is presented in Sect. $5$, together
with the transformation of symmetries. The algebra of symmetries is considered
in the next Section, where the appearance of two PST symmetries
from a single $\alpha$--symmetry transformation is demonstrated. 
In Section $6$ we summarise our results and make concluding remarks.
\section{Definitions}

We define the antisymmetric second rank tensor $A_{\mu \nu }$ in $d=5+1$ dimensions with
self--dual field strength $F_{\mu \nu \lambda }$ by the following relations: 
\begin{eqnarray}
A_{\mu \nu } &=&-A_{\nu \mu },\quad \mu ,\nu =0,1,2,3,4,5.  \label{1} \\
F_{\mu \nu \lambda } &=&\partial _{\mu }A_{\nu \lambda }+\partial _{\nu
}A_{\lambda \mu }+\partial _{\lambda }A_{\mu \nu },  \label{2} \\
F_{\mu \nu \lambda }^{\pm } &=&\frac{1}{2}\left( F_{\mu \nu \lambda }\pm 
\frac{1}{6}\varepsilon _{\mu \nu \lambda \sigma \rho \delta }F^{\sigma \rho
\delta }\right) ,  \label{3} \\
F_{\mu \nu \lambda }^{-} &=&0,  \label{4}
\end{eqnarray}

We use (anti)symmetrization with unit weight, for example:

\begin{eqnarray}
S_{(\mu \nu )} &=&\frac{1}{2}\left( S_{\mu \nu }+S_{\nu \mu }\right) ,
\label{5} \\
T_{[\mu \nu ]} &=&\frac{1}{2}\left( T_{\mu \nu }-T_{\nu \mu }\right) .
\label{6}
\end{eqnarray}

We can then prove the following identities valid for arbitrary
antisymmetric third rank tensors, a symmetric traceless tensor $\theta _{\mu
\nu }$ and an antisymmetric tensor $\Lambda _{\mu \nu }$:

\begin{eqnarray}
&&K_{\mu \nu \rho }^{\pm }H^{\pm \mu \nu \rho } = 0,\quad K_{\mu \nu \rho
}^{-}H^{+\lambda \nu \rho } + K_{\quad \nu \rho }^{-\lambda }H_{\mu }^{+\nu
\rho }=\frac{1}{3}\delta _{\mu }^{\lambda }K_{\alpha \nu \rho
}^{-}H^{+\alpha \nu \rho },  \label{7} \\
&&K_{\alpha [\nu \rho }^{\pm }\theta _{\lambda ]}^{\alpha } = \left(
K_{\alpha [\nu \rho }^{\pm }\theta _{\lambda ]}^{\alpha }\right) ^{\mp
},\quad K_{\mu \nu \rho }^{\pm }H^{\pm \lambda \nu \rho }-K_{\quad \nu \rho
}^{\pm \lambda }H_{\mu }^{\pm \nu \rho }=0,  \label{8} \\
&&K_{\alpha [\nu \rho }^{\pm }\Lambda _{\lambda ]}^{\alpha } = \left(
K_{\alpha [\nu \rho }^{\pm }\Lambda _{\lambda ]}^{\alpha }\right) ^{\pm
},\quad K_{\mu \nu \rho }^{\pm }H^{\pm \mu \lambda \sigma }+K_{\mu }^{\pm
\lambda \sigma }H_{\quad \nu \rho }^{\pm \mu }=  \nonumber \\
&&\qquad\qquad\qquad\qquad\qquad\qquad\qquad\qquad\qquad\qquad2\delta _{[\nu
}^{[\lambda }K_{\rho ]\alpha \beta }^{\pm }H^{\pm \sigma ]\alpha \beta },
\label{9} \\
&&K_{\mu \nu \rho }^{-}H^{+\mu \lambda \sigma }-K_{\mu }^{-\lambda \sigma
}H_{\quad \nu \rho }^{+\mu } = \frac{1}{3}K_{\alpha \beta \gamma
}^{-}H^{+\alpha \beta \gamma }\delta _{\nu }^{[\lambda }\delta _{\rho
}^{\sigma ]}  \nonumber \\
&&\qquad\qquad\qquad\qquad\qquad\qquad\qquad\qquad\qquad-2\delta _{[\nu
}^{[\lambda }K^{-\sigma ]\alpha \beta }H_{\rho ]\alpha \beta }^{+},
\label{10} \\
&&K^{-\mu \nu \rho }H_{\lambda \sigma \delta }^{+}+K_{\lambda \sigma \delta
}^{-}H^{+\mu \nu \rho }=\delta _{\lambda }^{[\mu }\delta _{\sigma }^{\nu
}\delta _{\delta }^{\rho ]}K_{\alpha \beta \gamma }^{-}H^{+\alpha \beta
\gamma }  \nonumber \\
&&\qquad\qquad\quad\qquad\qquad\qquad -9\delta _{[\lambda }^{[\mu }\delta
_{\sigma }^{\nu }K_{\delta ]\alpha \beta }^{-}H^{+\rho ]\alpha \beta
}+9\delta _{[\lambda }^{[\mu }K_{\sigma \delta ]\alpha }^{-}H^{+\nu \rho
]\alpha },  \label{11} \\
&&K^{\pm \mu \nu \rho }H_{\lambda \sigma \delta }^{\pm }-K_{\lambda \sigma
\delta }^{\pm }H^{\pm \mu \nu \rho } = 9\delta _{[\lambda }^{[\mu }\delta
_{\sigma }^{\nu }K_{\delta ]\alpha \beta }^{\pm}H^{\pm\rho ]\alpha \beta
}-9\delta _{[\lambda }^{[\mu }K_{\sigma \delta ]\alpha }^{\pm}H^{\pm\nu \rho
]\alpha }  \label{12}
\end{eqnarray}

\section{The $\alpha$--symmetry formalism for chiral\\ bosons}

We consider the following action, introduced in the work of Kavalov and
Mkrtchyan \cite{KM}, which involves tensor $A_{\mu \nu }$ and the sixth--rank
tensor $\theta _{\mu \nu \lambda ,\sigma \rho \delta }^{++}$, the latter being
self--dual over each set of three indices: 
\begin{equation}
S_{KM}=\int d^{6}x\left\{ -\frac{1}{6}F_{\mu \nu \lambda }F^{\mu \nu \lambda
}+\frac{1}{3}\theta _{\mu \nu \lambda ,\sigma \rho \delta }^{++}F^{-\mu \nu
\lambda }F^{-\sigma \rho \delta }\right\}   \label{13}
\end{equation}
Here 
\begin{eqnarray}
\theta _{\mu \nu \lambda ,\sigma \rho \delta } &=&\theta _{[\mu \nu \lambda
],[\sigma \rho \delta ]}=\theta _{\sigma \rho \delta ,\mu \nu \lambda },
\label{14} \\
\theta _{\mu \nu \lambda ,\sigma \rho \delta }^{++} &=&P_{\mu \nu \lambda
}^{+\quad \mu _{1}\nu _{1}\lambda _{1}}P_{\sigma \rho \delta }^{+\quad
\sigma _{1}\rho _{1}\delta _{1}}\theta _{\mu _{1}\nu _{1}\lambda _{1},\sigma
_{1}\rho _{1}\delta _{1}},  \label{15} \\
P_{\mu \nu \lambda }^{\pm \quad \mu _{1}\nu _{1}\lambda _{1}} &=&\frac{1}{2}%
\left( \delta _{\mu }^{\mu _{1}}\delta _{\nu }^{\nu _{1}}\delta _{\lambda
}^{\lambda _{1}}\pm \frac{1}{6}\varepsilon _{\mu \nu \lambda }^{\quad \mu
_{1}\nu _{1}\lambda _{1}}\right)   \label{16}
\end{eqnarray}
The tensor $\theta $ serves as a Lagrange multiplier. From the Lagrangian of $\left( \ref{13}\right)$  
we obtain the following equations of motion: 
\begin{eqnarray}
F_{\mu \nu \lambda }^{-}F_{\sigma \rho \delta }^{-} &=&0,  \label{17} \\
\partial _{\mu }\left( F^{-\mu \nu \lambda }-\theta ^{++\mu \nu \lambda
,\rho \sigma \delta }F_{\rho \sigma \delta }^{-}\right)  &=&0  \label{18}
\end{eqnarray}
These equations are equivalent to the self--duality-condition: 
\begin{equation}
F_{\mu \nu \lambda }^{-}=0  \label{19}
\end{equation}
and do not impose any restriction on the auxiliary field $\theta _{\mu \nu
\lambda ,\sigma \rho \delta }^{++}$. In addition to the usual gauge invariance: 
\begin{equation}
\delta A_{\mu \nu }=2\partial _{[\mu }\Lambda _{\nu ]}  \label{20}
\end{equation}
$S_{KM}$ is also invariant with respect to the so-called $\alpha$--symmetry with
vector parameter $\alpha ^{\rho }$ defined by: 
\begin{eqnarray}
\delta (\alpha )A_{\mu \nu } &=&\alpha ^{\rho }\left( F_{\rho \mu \nu
}^{-}+\theta _{\rho \mu \nu ,\lambda \sigma \delta }^{++}F^{-\lambda \sigma
\delta }\right)   \label{21} \\
\delta (\alpha )\theta _{\mu \nu \lambda }^{++\,\sigma \rho \delta }
&=&\left( P_{\theta }^{+}AP_{\theta }^{-}\right) _{\mu \nu \lambda
}^{\,\,\,\,\,\,\,\,\,\,\,\,\sigma \rho \delta }+\left( P_{\theta
}^{+}AP_{\theta }^{-}\right) _{\quad \mu \nu \lambda }^{\sigma \rho \delta
}-\alpha ^{\gamma }\partial _{\gamma }\theta _{\mu \nu \lambda }^{++\,\sigma
\rho \delta }  \label{22} \\
A &=&-\frac{3}{2}\delta _{\mu }^{\sigma }\delta _{\nu }^{\rho }\partial
_{\lambda }\alpha ^{\delta },\quad \left( P_{\theta }^{\pm }\right) _{\mu
\nu \lambda }^{\,\,\,\,\,\,\,\,\,\,\,\,\sigma \rho \delta }=P_{\mu \nu
\lambda }^{\pm \,\,\,\,\,\,\sigma \rho \delta }\mp \theta _{\mu \nu \lambda
}^{++\,\sigma \rho \delta }.  \label{23}
\end{eqnarray}
Using this approach for the Lagrangian of self--dual fields, in refs. \cite{KM}
the supersymmetric Lagrangians were constructed for all supergravities,
containing such fields in their supermultiplets (and even non--chiral
supergravities were presented in such a form). The $\alpha$--symmetry $\left( 
\ref{21}\right) $ is maintained in these Lagrangians, and plays an important
role in the closure of the algebra of symmetries of the theory. 
In particular the $\alpha$--part appears on the r.h.s of the commutator of 
two local supersymmetry transformations in all supergravities \cite{KM}: 
\begin{eqnarray}
\left[ \delta (\varepsilon _{2}),\delta (\varepsilon _{1})\right]  &=&\delta
(diff)+\delta (Lorentz)+\delta (gauge)+\delta (\alpha )  \label{24} \\
&+&\delta (susy)+(eq.\,\, of\,\, motion)  \nonumber
\end{eqnarray}
The explicit expressions depend, of course, on the specific theory
considered, but the algebra (\ref{24}) remains the same. Moreover, the parameter
of $\alpha$--symmetry is always equal to the parameter of the general coordinate
transformation: $\alpha ^{\mu }=\xi ^{\mu }$. As an example and fot further use,
we quote here some expressions for the simplest case of $d=6$, ${\cal N}=2$ chiral
supergravity (notation as in  \cite{KM}): 
\[
\alpha ^{\mu }=\xi ^{\mu }=\bar{\varepsilon}_{1}^{a}\gamma ^{\mu
}\varepsilon _{2a}
\]

The susy transformation of $\theta $ is:
\begin{equation}
\delta (\varepsilon )\theta _{\alpha\beta\gamma}^{++\,\alpha'\beta'\gamma'}=\left( P_{\theta
}^{+}EP_{\theta }^{-}\right) _{\alpha\beta\gamma}^{\,\,\,\,\,\,\,\,\,\,\,\,\alpha'\beta'\gamma'}+\left(
P_{\theta }^{+}EP_{\theta }^{-}\right) _{\,\,\,\,\,\,\,\,\quad\alpha\beta\gamma }^{\alpha'\beta'\gamma'}  \label{25.1}
\end{equation}

\begin{equation}
E=\frac{3}{2}\delta _{\alpha}^{\alpha'}\delta _{\beta}^{\beta'}e_{\gamma}^{\mu }\delta (\varepsilon
)e_{\mu }^{\gamma'}  \label{25.2}
\end{equation}

\section{The Pasti-Sorokin-Tonin formulation of the chiral boson}

The action constructed in ref. \cite{PST} is: 
\begin{equation}
S_{PST}=\int d^{6}x\left\{ -\frac{1}{6}F_{\mu \nu \lambda }F^{\mu \nu
\lambda }+\frac{2}{\left( \partial _{\rho }a\partial ^{\rho }a\right) }%
\partial ^{\mu }aF_{\mu \nu \lambda }^{-}F^{-\sigma \nu \lambda }\partial
_{\sigma }a\right\}  \label{26}
\end{equation}
where the scalar field $a$ is an auxiliary field, analogous to $\theta $ of
the previous Section. This action is invariant under the following local gauge
transformation: 
\begin{eqnarray}
\delta _{1}(\varphi )A_{\mu \nu } &=&\frac{2\varphi (x)}{\left( \partial
_{\rho }a\partial ^{\rho }a\right) }F_{\mu \nu \lambda }^{-}\partial
^{\lambda }a,\quad \delta _{1}a=\varphi (x)  \label{27} \\
\delta _{2}(\phi _{\nu })A_{\mu \nu } &=&\partial _{\mu }a\phi _{\nu
}-\partial _{\nu }\phi _{\mu },\quad \delta _{2}a=0  \label{28}
\end{eqnarray}
As in the previous Section, there is only one independent equation of motion,
in this case that of $A_{\mu \nu }$: 
\begin{equation}
\partial _{[\mu }\left( \frac{1}{\left( \partial _{\rho }a\partial ^{\rho
}a\right) }\partial _{\nu }aF_{\lambda \rho ]\sigma }^{-}\partial ^{\sigma
}a\right) =0  \label{29.1}
\end{equation}
with the general solution: 
\begin{equation}
F_{\mu \nu \lambda }^{-}\partial ^{\lambda }a=\delta _{2}(\phi _{\nu })\left[F_{\mu \nu \lambda}^{-}\partial ^{\lambda }a\right]  \label{30}
\end{equation}
which is equivalent to the self--duality condition due to the gauge invariance $%
\left(\ref{28}\right)$, which allows us to bring to zero the r.h.s of $\left(%
\ref{30}\right)$, and the self--duality equation follows. As mentioned above,
this approach was used for the construction of the $5$--brane action \cite{PST}, and
also the actions for chiral supergravities as in \cite{New2BSugra}.

In the next Section we shall establish a connection between the two approaches,
and in particular between expressions and symmetries of actions for chiral
supergravities.
\newpage
\section{The Ansatz}

Evidently actions $\left( \ref{13}\right) $ and $\left( \ref{26}\right) $
are connected by the ansatz for $\theta $: 
\begin{equation}
\theta _{\mu \nu \lambda }^{++\,\rho \sigma \delta }=6\left( \frac{\partial
_{\mu }a\partial ^{\rho }a}{\left( \partial _{\chi }a\partial ^{\chi
}a\right) }\delta _{\nu }^{\sigma }\delta _{\lambda }^{\delta }\right)
^{++}=6\left( N_{\mu \nu \lambda }^{\quad \rho \sigma \delta }\right) ^{++}
\label{31}
\end{equation}
\begin{equation}
\theta _{\mu \nu \lambda }^{++\,\rho \sigma \delta }F_{\rho \sigma \delta
}^{-}=6F_{\rho [\nu \lambda }^{-}\left( \frac{\partial _{\mu ]}a\partial
^{\rho }a}{\left( \partial _{\sigma }a\partial ^{\sigma }a\right) }-\frac{1}{%
6}\delta _{\mu ]}^{\sigma }\right)   \label{32}
\end{equation}
or, equivalently, using $\left( \ref{7}\right) $ to $\left( \ref{12}\right) $
and $\left( \ref{14}\right) $ to $\left( \ref{16}\right) $ 
\begin{eqnarray}
\theta _{\mu \nu \lambda }^{++\,\rho \sigma \delta } &=&6\left( \frac{%
\partial _{[\mu }a\partial ^{\gamma }a}{\left( \partial _{\chi }a\partial
^{\chi }a\right) }-\frac{1}{6}\delta _{[\mu }^{\gamma }\right) P_{\gamma \nu
\lambda ]}^{-\quad \rho \sigma \delta }  \nonumber \\
&=&3\left( P^{+}NP^{-}\right) _{\mu \nu \lambda }^{\quad \rho \sigma \delta
}+3\left( P^{+}NP^{-}\right) _{\quad \mu \nu \lambda }^{\rho \sigma \delta }
\label{33}
\end{eqnarray}
The ansatz, which we shall use to establish the connection between
supergravities, differs from (\ref{33}) in that all indices of $\theta $
have to be flat ones, so that world indices of derivatives of the scalar field
have to be transformed into flat ones by sixbeins. With this understanding,
it can  easily be checked term--by--term that Lagrangians of \cite{New2BSugra}
can be obtained from that of \cite{KM} by this substitution of $\theta $. The question arises:
What happens with the symmetry transformations and the algebra of symmetries? It is easy
to check, that the first PST symmetry $\left( \ref{27}\right)$ is a
particular case of the $\alpha$--symmetry $\left( \ref{21}\right) $, with the
following ansatz for the parameter $\alpha $: 
\begin{equation}
\alpha ^{\lambda }(x)=\frac{\varphi (x)\partial ^{\lambda }a}{\left(
\partial _{\chi }a\partial ^{\chi }a\right) }  \label{34}
\end{equation}
In particular, the entire transformation (\ref{22}) follows from the shift of $a
$ in the ansatz (\ref{33}). The same statement applies to the supersymmetry
transformation. The susy transformations of \cite{New2BSugra} are in
agreement with that of \cite{KM}, particularly with (\ref{25.1}), under the 
ansatz $\left( \ref{33}\right) $.
Since in the PST formalism the susy transformation of the auxiliary
scalar field $a$ is zero, the whole expression (\ref{25.1}) essentially
originates from the transformation of the metric in the scalar product of ansatz (%
\ref{33}), or from transformation of sixbeins in the same expression, needed for
conversion of world indices of $\partial _{\mu }a$ into flat ones of the tensor $%
\theta $. Next, there is no analogue for the second PST symmetry in ($\ref{28}$). 
This raises the question about the commutator of susy transformations ($\ref{24}$%
), which according to \cite{New2BSugra} contain both PST symmetries.
Remarkably, the expression for the $\alpha$--symmetry transformation in the
commutator $\left( \ref{24}\right) $ splits into a sum of two PST symmetries
with special parameters: 
\begin{eqnarray}
\delta (\alpha )A_{\mu \nu } &=&\frac{2(\alpha ^{\lambda }\partial _{\lambda
}a)}{\left( \partial _{\chi }a\partial ^{\chi }a\right) }F_{\mu \nu \rho
}^{-}\partial ^{\rho }a  \nonumber \\
&+&\partial _{\mu }a\left[ \frac{2F_{\nu \lambda \rho }^{-}\alpha ^{\lambda
}\partial ^{\rho }a}{\left( \partial _{\chi }a\partial ^{\chi }a\right) }%
\right] -\partial _{\nu }a\left[ \frac{2F_{\mu \lambda \rho }^{-}\alpha
^{\lambda }\partial ^{\rho }a}{\left( \partial _{\chi }a\partial ^{\chi
}a\right) }\right]   \label{35} \\
&=&\delta _{1}\left( \alpha ^{\lambda }\partial _{\lambda }a\right) A_{\mu
\nu }+\delta _{2}\left( \frac{2F_{\gamma \lambda \rho }^{-}\alpha ^{\lambda
}\partial ^{\rho }a}{\left( \partial _{\chi }a\partial ^{\chi }a\right) }%
\right) A_{\mu \nu }  \nonumber
\end{eqnarray}
Thus although the formalism \cite{KM} does not contain the second PST symmetry,
the latter arises, after substitution of the ansatz, from the $\alpha$--symmetry
transformation term in the commutator of the susy transformations.This last
point completes our establishment of the connection between supergravity Lagrangians of 
\cite{KM} and those of \cite{New2BSugra} (as communicated earlier privately \cite{MS}).

\section{Conclusions}

Above we have established the connection between two formalisms, 
i.e. that of $\alpha$--symmetry,
used about ten years ago for the construction of Lagrangians of
supergravities with (anti)--self--dual tensors, and the recently invented PST
formalism, which solves the same problem, and permits, in addition, the
construction of the $5$-brane Lagrangian. It appears, that Lagrangians of
supergravities are connected through the ansatz, which connects the
auxiliary field of both formalisms, i.e. the tensor $\theta $ and the scalar $a$. The
transformation of $\alpha$--symmetry terms into the PST symmetries, in the
algebra of supersymmetry transformations, has been demonstrated. Also the
agreement of rules for symmetries, particularly supersymmetries,
has been demonstrated above.

Beyond the scope of this connection there remains the problem of the Lagrangian
for the $5$--brane, which so far has been established in Lorentz invariant form
only in the PST formalism. Probably the latter can be
constructed within the $\alpha$--symmetry formalism by considering more general
expressions, satisfying the $\alpha$--symmetry requirement.
It may be noted that $\alpha$--symmetry strongly resembles reparametrization
invariance, and the problem is something like the construction of actions with
general coordinate invariance.

\textbf{Acknowledgments}

This work was supported in part by the U.S. Civilian Research and
Development Foundation under Award \#96-RP1-253 and by INTAS grants \#96-538,
\#93-1038 (ext).
R. Manvelyan is indebted to D.Sorokin for discussion and 
to the A. von Humboldt Foundation for financial support.


\end{document}